# Spatial Imaging of Magnetically Patterned Nuclear Spins in GaAs

J. Stephens, R.K. Kawakami, J. Berezovsky, M. Hanson, D.P. Shepherd, A.C. Gossard, D.D. Awschalom

*Center for Spintronics and Quantum Computation, University of California, Santa Barbara, CA 93106*

**Abstract**

We exploit ferromagnetic imprinting to create complex laterally defined regions of nuclear spin polarization in lithographically patterned MnAs/GaAs epilayers grown by molecular beam epitaxy (MBE). A time-resolved Kerr rotation microscope with ~1 micron spatial resolution uses electron spin precession to directly image the GaAs nuclear polarization. These measurements indicate that the polarization varies from a maximum under magnetic mesas to zero several microns from the mesa perimeter, resulting in large (~$10^4$ T/m) effective field gradients. The results reveal a flexible scheme for lateral engineering of spin-dependent energy landscapes in the solid state.

The ability to control nuclear and electron spin in solids is attracting increased interest for use in spintronics and quantum computation.[1,2] One possible route toward achieving such control is the recently discovered ferromagnetic imprinting of nuclear spins.[3] In this process, a ferromagnetic film induces large (up to ~20%) nuclear spin polarization in an adjacent GaAs epilayer under photoexcitation. This nuclear polarization in turn generates strong effective magnetic fields (up to ~0.9 T) that act on free-carrier electron spins through the contact hyperfine interaction. While ferromagnetic imprinting was previously studied in continuous films, the ability to imprint a complex, spatially varying magnetization onto the semiconductor nuclear spins via lithographic patterning of the adjacent ferromagnet may offer a route towards novel spin-based devices. In this case, laterally defined nuclear domains should result in spatially varying effective magnetic fields and field gradients, giving rise to local enhancement of the electron Larmor precession frequency and spin-dependent forces on electrons, respectively. The feasibility of such an approach, however, depends sensitively on electron and nuclear spin diffusion during the ferromagnetic imprinting process because these effects may limit the ability to precisely define nuclear domain structures as well as restrict the peak gradient by smearing the nuclear polarization.

In this paper we investigate how closely the nuclear polarization in the GaAs tracks the magnetization of patterned ferromagnet structures using a low temperature time-resolved Kerr microscope (TRKM) with ~1 µm resolution. We find that the spatial variation of the GaAs nuclear polarization approximates the shape of the patterned ferromagnet, with a transition from maximum to zero nuclear polarization occurring over a distance $\Lambda$ of ~10 µm, yielding effective magnetic field gradients of ~$10^4$ T/m at T =

5K. We also find that Λ remains roughly constant with increasing temperature, in contrast to the electronic mobility. This suggests that the longitudinal electron spin lifetime may play a role in setting the transition width. Finally, we observe a monotonic decrease in Λ with increasing GaAs epilayer thickness as well as fluctuations in the nuclear polarization as a function of lateral position under the ferromagnet.

Samples are grown by MBE and have the following structure: Type-A MnAs(15 or 25 nm)/n-GaAs(100 nm, 250 nm, 500 nm)/Al$_{0.75}$Ga$_{0.25}$As(400 nm)/n$^+$-GaAs(100) substrate. All layers up to the MnAs are grown in a separate chamber to maximize the optical quality of the n-GaAs layer (Si: $7\times10^{16}$cm$^{-3}$) and are capped with arsenic to protect the films during transfer between the two chambers.[4] Several characterization measurements are performed on the patterned films to ensure that neither sidewall roughness nor non-uniform magnetization of the MnAs structures limits the abruptness of changes in nuclear polarization. Figure 1(a) shows an optical micrograph of an array of 100 μm x 100 μm square MnAs mesas defined by optical lithography followed by a chemically selective wet etch (1 minute HCl:H$_2$O,1:10). Comparison of hysteresis loops for a patterned and an unpatterned sample taken at T = 5 K in a commercial superconducting quantum interference device magnetometer (SQUID) (Fig. 1(b)) reflect unchanged bulk magnetic properties. Magnetic force microscopy (MFM) of a magnetized MnAs mesa indicates that a single magnetic domain extends to the mesa edges in zero applied field as shown in Fig. 1(c), while atomic force microscopy (AFM) (not shown) identifies mesa sidewall widths of ~100 nm, lateral edge roughness of < 1 μm, and MnAs root mean square vertical roughness (RMS) of ~0.4 nm. Samples are mounted face down onto c-axis sapphire using a transparent epoxy and the substrate is removed up to the 400 nm AlGaAs layer using a chemically selective etch.

Nuclear polarization in the GaAs layer is determined by measuring the Larmor precession frequency of electron spins via time-resolved Kerr rotation.[5,6] The measurement is performed with the TRKM[7] (Fig. 2(a)) which is used to image electron spin dynamics in GaAs with ~1 μm resolution from T = 5 K to room temperature. Pump and probe pulses from a Ti:sapphire laser (~150 fs pulse width, 76 MHz repetition rate, wavelength ~815 nm) are focused through a microscope objective (numerical aperture = 0.73) which can be scanned laterally with ~20 nm precision. The circularly polarized pump pulse (~450 μW) incident normal to the sample excites spin-polarized electrons in the GaAs layer in the direction of its propagation. After a time delay $\Delta t$, a linearly polarized probe pulse (~350 μW) passes through this non-equilibrium spin population and experiences a Kerr rotation $\theta_K$ proportional to the component of net spin normal to the GaAs layer. The temporal evolution of this net spin polarization is obtained by measuring $\theta_K$ while scanning $\Delta t$. In the presence of a transverse magnetic field, the electron spins precess coherently in a plane perpendicular to the field at the Larmor frequency $\nu_L$. Here, $\nu_L$ is proportional to the total effective transverse field comprised of the applied field ($B_{app}$, set to 0.21 T) and the nuclear contact hyperfine field ($B_N$), and is given by $\nu_L = g\mu_B(B_{app} + B_N)/h$ where $\mu_B$ is the Bohr magneton and $g$ = -0.44 for GaAs. In general, the Larmor frequency may be modulated by other effects (e.g. fringing fields), however as discussed below, they contribute negligibly to $\nu_L$ in these experiments. In our experiment the Larmor precession manifests as an exponentially damped cosine, $\theta_K(\Delta t) = A \exp(-\Delta t/T_2^*) \cos(2\pi\nu_L\Delta t)$ where $T_2^*$ is the effective transverse spin lifetime. Thus, fitting $\theta_K$ as a function of $\Delta t$ yields a measure of $B_N$, which in turn is proportional to the nuclear polarization.[8,9,10] In addition to the pump and probe beams, a third linearly polarized beam (813nm, 30mW continuous wave diode laser) is incident to the sample through the sapphire and serves to excite the photocarriers in the GaAs necessary to establish the nuclear imprint.[3,11] This "imprinting" beam is defocused to a spot size of ~500 μm to provide broad, slowly varying illumination over the scan area. Figure 2(b) shows a delay scan taken at the center of a MnAs mesa as well as one taken on the bare GaAs between two mesas. The higher precession frequency on the mesa indicates the presence of non-zero nuclear polarization corresponding to $B_N$ of ~0.2 T.

Figure 2(c) shows a charge coupled device image of the tightly focused pump and probe beams as well as the centermost region of the large imprinting beam on a MnAs mesa which provides the background illumination. Because we are interested in the nuclear polarization profile created by the broad imprinting beam, we ensure that the pump and probe beams do not significantly alter the nuclear polarization by studying the laboratory time dependence of the nuclear polarization at the center of a MnAs mesa (not shown). First, the pump-probe beams are blocked for ~15 minutes in order to allow the nuclear polarization to reach steady-state conditions under illumination from the imprinting beam only. Then the pump-probe beams are unblocked and $\nu_L$ is measured as a function of lab time. We observe a change in $\nu_L$ of < 5% indicating that the effect of the pump-probe beams is negligible compared to the imprinting beam. In addition, we note that because the imprinting beam is linearly polarized, nuclear polarization induced by optically driven dynamic nuclear polarization is negligible.[9,12]

In order to quantify how precisely the nuclear polarization tracks the shape of the MnAs mesas, we construct two dimensional images of $B_N$ by scanning the microscope objective relative to the fixed sample and performing a time delay scan at each position. Figures 3(a) and 3(b) depict 150 μm x 150 μm scans (3 μm step size) of $B_N$ over a MnAs mesa for samples of GaAs epilayer thickness of 100 nm and 500 nm, respectively. Qualitatively, these plots show a transition from maximum nuclear polarization under the MnAs mesas to zero nuclear polarization in regions where the MnAs has been etched away. Analysis of individual line scans over the mesa edge (Fig. 3(c)) yields the transition width, Λ, which is operationally defined as the distance over which $B_N$ increases from 10% to 90% of its average value on the mesa, $<B_N>$. Comparing the two samples, Λ is larger for the sample with 500 nm n-GaAs. Furthermore, for the 100 nm sample the effective field $<B_N>$ is 0.34 T (6% polarization) and the maximum gradient is $4.4 \times 10^4$ T/m, while these values are 0.28 T (5% polarization) and $1.8 \times 10^4$ T/m for the 500 nm sample.

Another striking feature of the data is the appearance of fluctuations in the magnitude of nuclear polarization as a function of lateral position on the sample, which do not correlate to any morphological features measured via AFM. These fluctuations in the nuclear polarization are more evident in the 100 nm sample (Fig. 3(a), left panel) than in the 500 nm sample. To determine whether they are associated with inhomogeneity in the imprinting beam or are associated with sample inhomogeneity, the imprinting beam is translated laterally and the nuclear polarization is imaged a second time. We find that the structure in the nuclear polarization is independent of imprinting beam position, indicating that the fluctuations are due to some sample inhomogeneity. Additional measurements performed on unpatterned samples show similar behavior indicating that the fluctuations are not an artifact of the patterning process and that the decrease in nuclear polarization at the edge of a mesa is not due to spatial variation of the imprinting beam intensity. Finally, Fig. 3(d) demonstrates that spreading of the nuclear polarization does not preclude the ability to define more complex nuclear domains.

The value of the width over which the nuclear polarization decreases, averaged over the mesa edges (Fig. 4(a), filled squares), yields values of $\Lambda = 9$ µm, 13 µm, and 17 µm for GaAs thicknesses of 100 nm, 250 nm, and 500 nm, respectively, at 5 K. As a measure of the fluctuations in $B_N$ we calculate the standard deviation ($\sigma_N$) divided by the mean ($<B_N>$) for the center 60 µm x 60 µm region of the 100 µm x 100 µm mesas. These data (Fig. 4(b), filled squares) show that the magnitude of the fluctuations decreases markedly with increasing GaAs thickness. We verify these trends in $\Lambda$ and $\sigma_N/<B_N>$ as a function of GaAs thickness by fabricating a second sample set (with identical GaAs thicknesses) from a new set of wafers. The results for $\Lambda$ and $\sigma_N/<B_N>$ are summarized by the open squares in Figs. 4(a) and 4(b), respectively. The agreement between the two sample sets indicates these quantities are robust even in the presence of growth-to-growth variations in the base material.[13] We also note that this consistency in $\Lambda$ and $\sigma_N/<B_N>$ remains despite the presence of sample-to-sample variation in the magnitude of $B_N$.

To investigate the origin of the fluctuations in nuclear polarization, we measure the nuclear polarization for a 100 nm GaAs layer covered by 12 nm of MBE-deposited Fe (open circle in Fig. 4(b)).[14] We observe that $\sigma_N/<B_N>$ for the Fe sample is significantly lower than that of a MnAs sample with equal GaAs thickness. This seems to indicate that the fluctuations are not driven by processes in the GaAs layer, but rather stem from variation inherent to the ferromagnet or ferromagnet/GaAs interface. We note that the lattice mismatch of the MnAs/GaAs system is larger than that of Fe/GaAs and that the nucleation processes during the growth of these two materials differ significantly.[15,16,17] In particular, observation of the reflection high energy electron diffraction pattern reveals that our MnAs films evolve from type-B to type-A material during the first few nanometers of growth. These differences may result in greater local variation of quantities such as the Schottky barrier in the GaAs/MnAs structures. This in turn may result in spatial variation in the efficiency of the ferromagnetic proximity polarization (FPP) of electron spin, which is known to be an underlying mechanism driving the nuclear polarization.[11] In addition, scans of MnAs samples at T = 60 K (at which there is negligible nuclear polarization) show spatially uniform $\nu_L$, indicating that the nuclear polarization fluctuations observed at T = 5 K are not due to inhomogeneous strain-induced changes in the Landé g-factor. These high temperature measurements also place an upper bound on contributions to $\nu_L$ from sources other than the applied and nuclear fields. For instance, one may reasonably expect that fringing fields (since the Curie temperature[16] of MnAs is 313 K) or Rashba splitting (resulting from drift due to strain-induced inhomogeneous band bending) would result in some modification of $\nu_L$ near the mesa edges. While these effects may indeed be present, they are not large enough to detect.

Finally, we investigate whether the electronic mobility is involved in determining $\Lambda$. Since photoexcited electrons polarized via the FPP process are necessary to generate the nuclear polarization, the magnitude of $\Lambda$ is likely related to the motion of these carriers (via drift or diffusion). Because the diffusion length and drift velocity both depend on electron mobility μ ($\ell_{diff} \sim \sqrt{\mu}$, $v_{drift} \sim \mu$), we investigate the role of electron motion by comparing the temperature dependence of μ and Λ for one of the 500 nm samples. Measurements from T = 5-35 K (Fig. 4(c) filled squares) show that Λ remains approximately constant as a function of temperature.[18] Mobility measurements are performed on Hall bars[19] at similar temperatures and under approximately the same illumination intensity as that provided by the imprinting beam (Fig. 4(c) filled circles). We find that μ increases with temperature, from a value of 3400 cm$^2$/V-s at T = 5 K to 5100 cm$^2$/V-s at T = 35 K. This discrepancy in the temperature dependences of Λ and μ indicates that a simple model explaining Λ in terms of only the electron drift and diffusion is inadequate. Other factors such as the electron spin lifetime, enhanced electron spin diffusion[20,21] and interfacial reflection of electrons[22] must be considered. In particular, one explanation for the different temperature dependences of Λ and mobility may be related to a decrease of the longitudinal electron spin lifetime $T_1$ with increasing temperature.[23] Given the temperature dependence of the mobility, we calculate an approximate $T_1$ by fitting linecuts of nuclear polarization with a simple model that estimates the steady-state diffusion profile of electron spin emanating from beneath the MnAs mesas.[24] Figure 4(d) shows an order of magnitude decrease (50 ns to 5 ns) in $T_1$ as calculated from the model from T = 5-35 K, as well as a representative linecut and fit of $B_N$ (inset). Another potential factor in determining Λ is the diffusion of nuclear spin, but to date we have not observed propagation of nuclear domain boundaries with time scales characteristic of nuclear diffusion (~minutes).[25] Further experimental and theoretical studies are needed to investigate how these factors affect Λ and its dependence on temperature and GaAs thickness.

In conclusion, we have demonstrated the patterning and imaging of spin-polarized nuclear domains in GaAs epilayers. Through the contact hyperfine interaction, these nuclear domains act on electron spins as localized magnetic fields with steep magnetic field gradients, the width of which depend on the thickness of the GaAs epilayer. This ability to pattern the nuclear polarization and its resulting effective fields and field gradients offers a flexible, scalable technique to spatially tailor a spin-dependent energy landscape, which in turn could lead to new concepts for spintronic devices as well as provide an intriguing tool for mesoscopic spin manipulation.

The authors wish to thank R. J. Epstein and D. K. Young for technical assistance and E. Johnston-Halperin and Y. Kato for stimulating discussions. We acknowledge support from DARPA/ONR N00014-99-1-1096, AFOSR F49620-02-10036, ARO DAAD19-01-1-0541 and NSF DMR-0071888.

**References**

[1] D. Loss and D. P. DiVincenzo, Phys. Rev. A **57**, 120 (1998).

[2] B. E. Kane, et al., Nature **393**, 133 (1998).

[3] R. K. Kawakami et al., Science **294**, 131 (2001).

[4] Growth conditions for MnAs: T = 240 C; growth rate ~0.4 nm/min; $As_4$/Mn beam flux ratio ~400.

[5] J. M. Kikkawa et al., Science **277**, 1284 (1997).

[6] S. A. Crooker, et al., Phys. Rev. Lett. **77**, 02814 (1997).

[7] W. K. Hiebert, A. Stankiewicz, and M. R. Freeman, Phys. Rev. Lett. **79**, 1134 (1997).

[8] D. Paget, G. Lampel, and B. Sapoval, Phys. Rev. B **15**, 5780 (1977).

[9] J. M. Kikkawa, D. D. Awschalom, Science **287**, 473 (2000).

[10] G. Salis et al., Phys. Rev. Lett. **86**, 2677 (2001).

[11] R. J. Epstein, et al., Phys. Rev. B **65**, 121202(R) (2002).

[12] Optically induced DNP is not strictly zero due to the presence of an applied field. The level of nuclear polarization due to this process with our modest field (~0.2 T), however, is negligible, as shown in Ref. 9.

[13] Of the two 500 nm samples shown in Figs. 3 and 4, one had 15 nm of MnAs while the other had 25 nm. The fact that the values of $\Lambda$ and the $\sigma_N/<B_N$ agree closely for the two samples suggests that these quantities are sensitive to the GaAs/MnAs interface.

[14] Fe grown at room temperature, growth rate ~0.1 nm/min.

[15] G. A. Prinz and J. J. Krebs, Appl. Phys. Lett. **39**, 397 (1981).

[16] M. Tanaka, Semicond. Sci. Technol. **17** 327 (2002).

[17] F. Schippan, et al., Appl. Phys. Lett. **76** 834 (2000).

[18] We note that although there is some mesa-to-mesa and sample-to-sample variation in the temperature dependence of $\Lambda$, all measurements have shown flat or slightly decreasing $\Lambda$ with temperature.

[19] Hall bars with indium contacts were patterned from the n-GaAs epilayer after removal of MnAs and were driven with 1 µA current. Under illumination, a carrier density of n = 6e15 cm$^{-3}$ was obtained.

[20] J. M. Kikkawa and D. D. Awschalom, Phys. Rev. Lett. **80**, 4313 (1998).

[21] M. E. Flatté and J. M. Byers, Phys. Rev. Lett. **84**, 4220 (2000).

[22] L. Sham, et al., Phys. Rev. Lett. **89**, 156601 (2002).

[23] P. H. Song and K.W. Kim, Physical Review B **66**, 035207 (2002).

[24] Our model assumes that the MnAs mesa acts as a step function source *S(x)* of electron spin, which results in a spin distribution *F(x, t)* given by $\partial F/\partial t = D_e \partial^2 F/\partial x^2 - F/T_1 + S(x)$. By modeling our data with the steady-state spin distribution we obtain a value for the product $D_e T_1$ where $D_e$ is the electron diffusivity. We can thus calculate "$T_1$" using the Einstein relation between mobility and diffusivity.

[25] D. Paget, Phys. Rev. B **25**, 4444 (1982).

**Figure Captions**

Fig. 1. (a) Optical micrograph of MnAs mesas. (b) Hysteresis loops taken at 5 K with field // GaAs[110] in a commercial SQUID for patterned (top) and unpatterned (bottom) MnAs epilayers. (c) Two different MFM images showing phase ($\Delta\phi$) contrast between the left and right edges of the same MnAs mesa indicating a single magnetic domain magnetized in plane.

Fig. 2. (a) Schematic of the experimental geometry. (b) Larmor precession of electron spins underneath a MnAs mesa (bottom) and between mesas (top). (c) CCD image of a MnAs mesa taken through the microscope objective showing the small overlapped pump and probe spots.

Fig. 3. (a,b) Spatial images of ferromagnetically-imprinted nuclear domains in 100 and 500 nm thick n-GaAs epilayers, respectively, measured at 5K and 0.21T applied field. The scale in the surface plots is the same in each case, while in the color map plots it is optimized for each sample. (c) A line scan on the 500 nm GaAs sample (0.5 µm step size) showing the definition of $\Lambda$ as 10% to 90% of $<B_N>$. (d) Spatial image showing a more complex nuclear domain pattern.

Fig. 4. (a) Nuclear polarization transition width $\Lambda$ as a function of GaAs thickness for two sample sets grown on different substrates. (b) Fluctuation in nuclear polarization $\sigma_N/<B_N>$ as a function of epilayer thickness for two sample sets, where $\sigma_N$ is the standard deviation and $<B_N>$ is the mean of $B_N$ under a MnAs mesa. Also shown (open circle) is one measurement on a sample with Fe in place of MnAs. (c) Transition width (squares)

and electronic mobility (circles) as a function of temperature on the 500 nm sample. (d) Fitted values of $T_1$ based on measured linecuts of nuclear polarization as a function of temperature. Inset: measured linecut of nuclear polarization (filled circles) and fit (solid line) of nuclear polarization that yields $T_1$ at T = 5 K for the 500 nm thick n-GaAs epilayer.

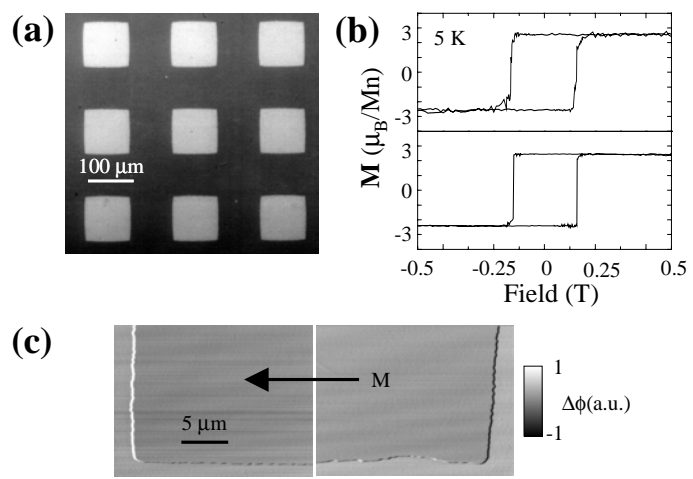

Figure 1

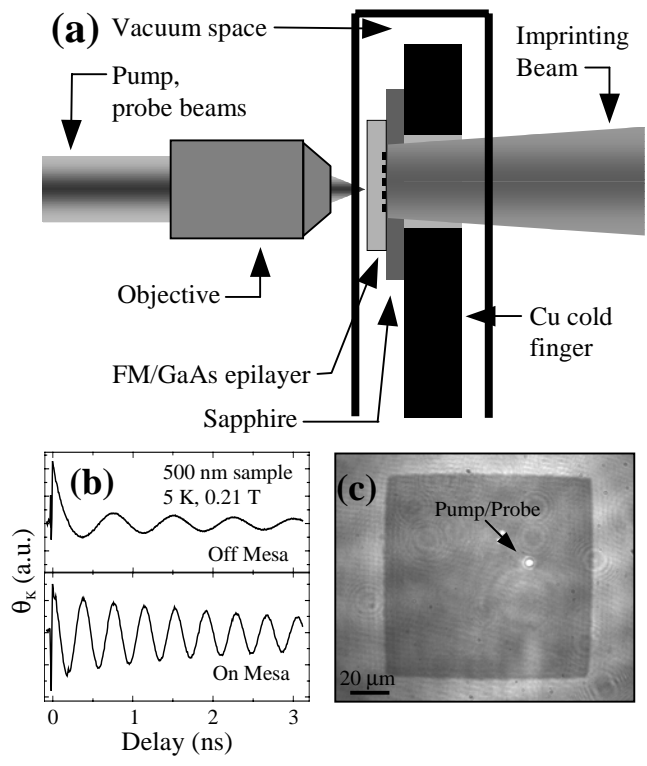

Figure 2

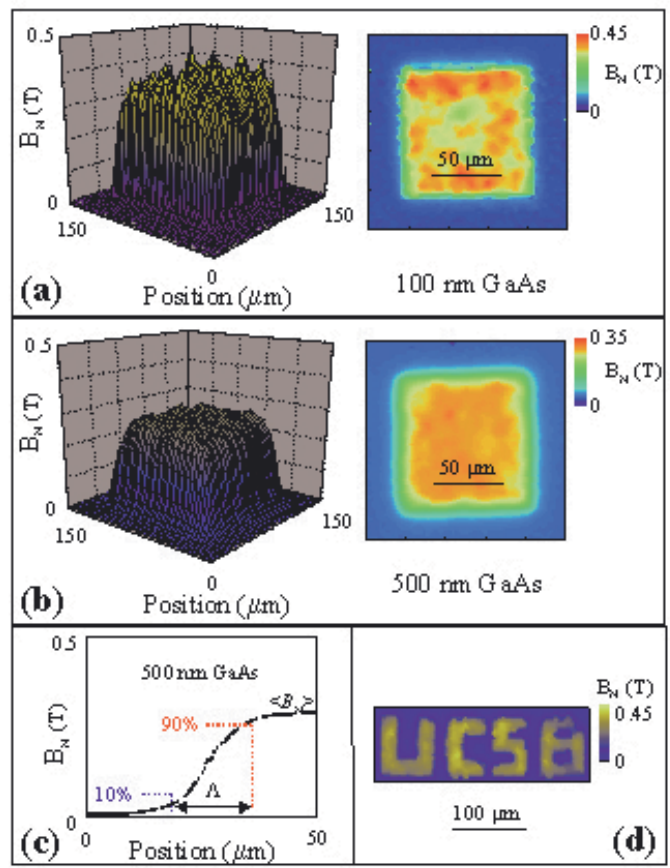

Figure 3

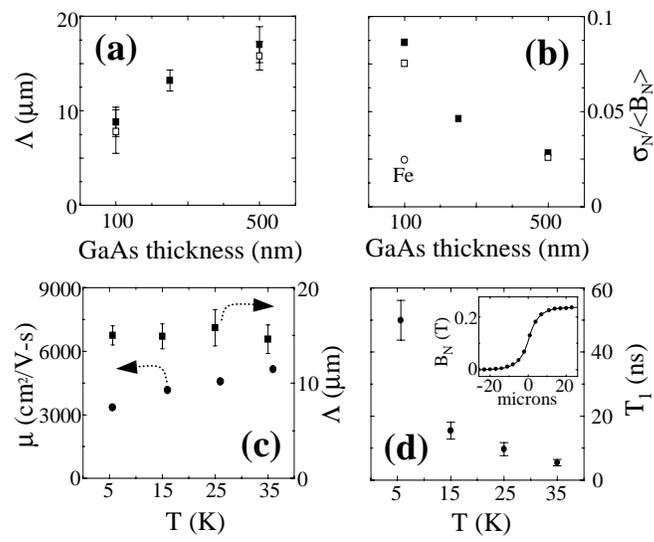

Figure 4